\documentclass{aa}
\usepackage{graphicx}
\usepackage{longtable}
\usepackage{natbib}
\usepackage{url}
\bibpunct{(}{)}{;}{a}{}{,}
\usepackage{txfonts}
\newcounter{Rco}
\newcommand{\Ionst}[1]{\setcounter{Rco}{#1}\Roman{Rco}}
\newcommand{\Ion}[2]{\mbox{#1\,{\scriptsize\Ionst{#2}}}}
\newcommand{\Ionw}[3]{\mbox{#1\,{\scriptsize\Ionst{#2}}~$\lambda\,#3$\,\AA}}

\newcommand{\Ionww}[3]{\mbox{#1\,{\scriptsize\Ionst{#2}}~$\lambda\lambda\,#3$\,\AA}}

\newcommand{\logg}{\mbox{$\log g$}}
\newcommand{\loggw}[1]{\mbox{$\log g\hspace{-0.5mm} =\hspace{-0.5mm}  #1$}}

\newcommand{\sA}[1]{\mbox{(Fig.\,\ref{#1})}}

\newcommand{\se}[1]{\mbox{Sect.\,\ref{#1}}}

\newcommand{\sK}[1]{\mbox{(Sect.\,\ref{#1})}}

\newcommand{\Teff}{\mbox{$T_\mathrm{eff}$}}
\newcommand{\Teffw}[1]{\mbox{$\Teff\hspace{-0.5mm} =\hspace{-0.5mm} #1 \,\mathrm{kK}$}}

\newcommand{\re}{\object{RE\,0503$-$289}}
\begin{document}
\title{Stellar laboratories:
       new Ge\,V and Ge\,VI oscillator strengths \\ and their validation in the hot white dwarf \re
        \thanks
        {Based on observations made with the NASA-CNES-CSA Far Ultraviolet Spectroscopic Explorer.
        }
        \thanks
        {Tables 2 and 4 are available in electronic form
         at the CDS via anonymous ftp to cdsarc.u-strasbg.fr (130.79.128.5)
         or via http://cdsweb.u-strasbg.fr/cgi-bin/qcat?J/A+A/
        }
      }
\titlerunning{Stellar laboratories: new Ge\,V and Ge\,VI oscillator strengths}

\author{T\@. Rauch\inst{1}
        \and
        K\@. Werner\inst{1}
        \and
        \'E\@. Bi\'emont\inst{2,3}
        \and
        P\@. Quinet\inst{2,3}
        \and
        J\@. W\@. Kruk\inst{4}
        }

\institute{Institute for Astronomy and Astrophysics,
           Kepler Center for Astro and Particle Physics,
           Eberhard Karls University,
           Sand 1,
           72076 T\"ubingen,
           Germany,
           \email{rauch@astro.uni-tuebingen.de}
           \and
           Astrophysique et Spectroscopie, Universit\'e de Mons -- UMONS, 7000 Mons, Belgium
           \and
           IPNAS, Universit\'e de Li\`ege, Sart Tilman, 4000 Li\`ege, Belgium
           \and
           NASA Goddard Space Flight Center, Greenbelt, MD\,20771, USA}

\date{Received 15 July 2012; accepted 14 August 2012}

\abstract {State-of-the-art spectral analysis of hot stars by means of non-LTE model-atmosphere
           techniques has arrived at a high level of sophistication. The analysis of
           high-resolution and high-S/N spectra, however, is strongly restricted by the
           lack of reliable atomic data for highly ionized species from intermediate-mass
           metals to trans-iron elements. Especially data for the latter has only been sparsely
           calculated. Many of their lines are identified in spectra of extremely hot,
           hydrogen-deficient post-AGB stars. A reliable determination
           of their abundances establishes crucial constraints for AGB nucleosynthesis
           simulations and, thus, for stellar evolutionary theory.
          }
          {In a previous analysis of the UV spectrum of \re,
           spectral lines of highly ionized Ga, Ge, As, Se, Kr, Mo, Sn, Te, I, and Xe
           were identified. Individual abundance determinations are hampered
           by the lack of reliable oscillator strengths. Most of these
           identified lines stem from \Ion{Ge}{5}.
           In addition, we identified \Ion{Ge}{6} lines for the first time.
           We calculated \Ion{Ge}{5} and \Ion{Ge}{6}
           oscillator strengths in order to reproduce the observed spectrum.
          }
          {We newly calculated \Ion{Ge}{5} and \Ion{Ge}{6} oscillator strengths
           to consider their radiative and collisional bound-bound transitions
           in detail in our non-LTE stellar-atmosphere models
           for the analysis of the Ge\,{{\sc iv}} -- {{\sc vi}} spectrum exhibited in
           high-resolution and
           high-S/N
           FUV (\emph{FUSE}) and
           UV (ORFEUS/\emph{BEFS}, \emph{IUE}) observations of \re.
          }
          {In the UV spectrum of \re, we identify
            four \Ion{Ge}{4},
              37 \Ion{Ge}{5}, and
           seven \Ion{Ge}{6} lines.
           Most of these lines are identified for the first time in any star.
           We can reproduce almost all \Ion{Ge}{4}, \Ion{Ge}{5}, and \Ion{Ge}{6} lines in the
           observed spectrum of \re\ (\Teffw{70}, \loggw{7.5}) at $\log \mathrm{Ge} = -3.8  \pm 0.3$
           (mass fraction, about 650 times solar). The Ge\,{{\sc iv}} / {{\sc v}} / {{\sc vi}}
           ionization equilibrium, that is a very sensitive \Teff\ indicator, is 
           reproduced well.
          }
          {Reliable measurements and calculations of atomic data are a prerequisite for
           stellar-atmosphere modeling. Our oscillator-strength calculations have allowed,
           for the first time, \Ion{Ge}{5} and \Ion{Ge}{6} lines to be successfully reproduced
           in a white dwarf's (\re) spectrum and to determine its photospheric Ge abundance.
          }


\maketitle

\section{Introduction}
\label{sect:intro}

Any model-atmosphere calculation is strongly dependent on the
available and reliable atomic data, which is a crucial input.
Especially for highly ionized species and
higher atomic mass, published data becomes rather sparse.
A close inspection of the UV spectrum of the hot white dwarf \re\ by
\citet{werneretal2012} has shown that a large number of the
hitherto unidentified observed spectral lines stem from
trans-iron elements, namely Ga, Ge, As, Se, Mo, Sn, Te, and I.

Identification of the respective spectral lines is fairly straightforward because
atomic databases like
NIST\footnote{\url{http://www.nist.gov/pml/data/asd.cfm}} and
Kelly's database\footnote{\url{http://www.cfa.harvard.edu/ampcgi/kelly.pl}}
have partly included the strongest lines of these elements with accurate wavelengths,
whereas a quantitative analysis requires adequate spectral modeling.
This is hampered by the fact that line strengths for trans-iron elements, when available at all, 
are relative intensities measured from emission line spectra.
Exploratory atmosphere models that are based on the LTE assumption
to calculate occupation numbers of the atomic levels of an ion and on $\log gf$ values
scaled to match the relative line strengths may show that the line identifications are correct.
A reliable abundance analysis, however, is impossible owing to the lack of measured or calculated
transition probabilities.

The line identification was demonstrated by \citet{werneretal2012} in the case of \re.
It is a hot (\Teffw{70}, \loggw{7.5}), helium-rich DO-type white dwarf
(\object{WD\,0501$-$289},
$\alpha_\mathrm{2000}=05^\mathrm{h}03^\mathrm{m}55\fs 513$, $\delta_\mathrm{2000}=-28\degr 54\arcmin 34\farcs 57$),
which is well-suited to UV spectroscopy because its spectrum is
only slightly contaminated by interstellar absorption.
\citet{werneretal2012} performed an abundance analysis of
Kr and Xe where level energies and oscillator strengths of \Ion{Kr}{6},
\Ion{Kr}{7}, \Ion{Xe}{6}, and \Ion{Xe}{7} were already published.
They determined
$\log \mathrm{Kr} = -4.3 \pm 0.5$ and
$\log \mathrm{Xe} = -4.2 \pm 0.6$
(mass fractions) and
identified a variety of lines of the other trans-iron elements
mentioned above.

\begin{figure}[ht!]
   \resizebox{\hsize}{!}{\includegraphics{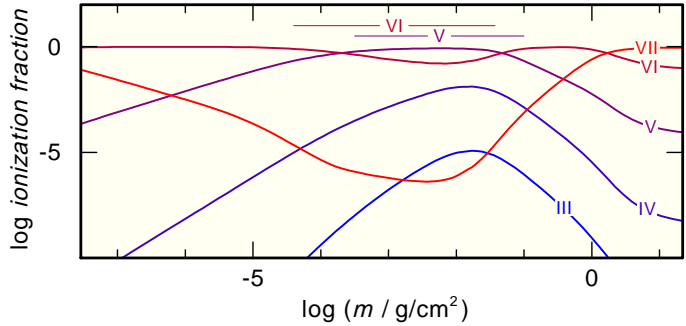}}
    \caption{Ionization fractions of Ge\,{\sc iii} - {\sc vii}.
             The formation depths of the \Ion{Ge}{5} and \Ion{Ge}{6}
             line cores are marked at the top.
            }
   \label{fig:ionization}
\end{figure}

Only level energies and (partly) relative line strengths
were accessible for Ge. A test calculation of an H+Ge-composed model atmosphere with the
relevant parameters (\Teffw{70}, \loggw{7.5}), $\log \mathrm{Ge} = -4$)
shows that \Ion{Ge}{5} and \Ion{Ge}{6} are dominant in the
line-forming region (Fig.~\ref{fig:ionization}).
Consequently, we calculated transition probabilities anew for
\Ion{Ge}{5} and \Ion{Ge}{6} \sK{sect:getrans}.
In \se{sect:observation}, we briefly introduce the available observed
spectra, which are used for our Ge abundance analysis of \re\
that is presented in \se{sect:model}.
In \se{sect:tefflogg} we re-assess the effective temperature
of \re\ based on the \Ion{C}{3} / \Ion{C}{4} ionization balance.
Results and conclusions are summarized in \se{sect:results}.

\section{Transition probabilities in \Ion{Ge}{5} and \Ion{Ge}{6}}
\label{sect:getrans}
There are not very many
transition probabilities or oscillator strengths in \Ion{Ge}{5} and \Ion{Ge}{6} ions. 
In \Ion{Ge}{5}, some pioneering 
HFR\footnote{Hartree-Fock with relativistic corrections} and 
MCDF\footnote{Multi-Configuration Dirac-Fock} \citep{grantmckenzie1980,grantetal1980}
results were reported by \citet{quinetbiemont1990,quinetbiemont1991}
but the work of these authors was limited to 3d-4p and 3d-4f transitions in nickel-like ions (\Ion{Ge}{5} - \Ion{Pb}{55}). 
More recent work comes from Safronova and co-workers 
\citep{safronovaetal2000,hamshaetal2004,safronovaetal2006a,safronovaetal2006b,safronovasafronova2006},
who performed relativistic many-body calculations for multipole transitions (E1, M1, E2, M2, E3, M3) originating in the ground states.

In \Ion{Ge}{6}, the available results are limited to forbidden transitions in 3d and 3d$^9$ configurations 
\citep{biemonthansen1989} and to the theoretical investigation of electric dipole transitions between 
3d$^9$ and d$^8$p configurations in  zinc, gallium, and germanium ions \citep{jucysetal1968}.

As there is no uniform set of oscillator strengths available for all the transitions of Ge ions observed in the present work, 
we decided to perform the relevant calculations. The method adopted here is the relativistic Hartree-Fock approach 
frequently referred to in the literature as the HFR or Cowan's method \citep{cowan1981}.

For \Ion{Ge}{5}, configuration interaction has been considered among the configurations 
3d$^{10}$, 3d$^9$$n$s ($n$ = 4--7), 3d$^9$$n$d ($n$ = 4--7), 3d$^8$4s$^2$, 3d$^8$4p$^2$, 3d$^8$4d$^2$,
3d$^8$4f$^2$, 3d$^8$4s$n$s ($n$ = 5--7), 3d$^8$4s$n$d ($n$ = 4--7), and 3d$^8$4p4f for the even parity,
and
3d$^9$$n$p ($n$ = 4--7), 3d$^9$$n$f ($n$ = 4--7), 3d$^8$4s$n$p ($n$ = 4--7), 3d$^8$4s$n$f ($n$ = 4--7), and 3d$^8$4p4d 
for the odd parity.
Using experimental energy levels reported by \citet{sugarmusgrove1993} and \citet{churilovetal1997}, the
radial integrals (average energies, Slater integrals, spin-orbit parameters) of 3d$^{10}$, 3d$^9$$n$s
($n$ = 4--7), 3d$^9$$n$p ($n$ = 4--6), 3d$^9$$n$d ($n$ = 4,5), 3d$^9$4f and 3d$^8$4s4p were optimized by
a well-established least-squares fitting procedure. In this process, the 3d$^9$4d $^1$S$_0$ level at
493765.5 cm$^{-1}$ \citep{sugarmusgrove1993} and the 3d$^8$4s4p ($J$=1) level at 673405 cm$^{-1}$
\citep{churilovetal1997}, affected by larger uncertainties, were not considered.

For \Ion{Ge}{6}, the configurations included in the HFR model were 3d$^9$, 3d$^8$4s, 3d$^8$5s, 3d$^8$4d,
3d$^8$5d, 3d$^7$4s$^2$, 3d$^7$4p$^2$, 3d$^7$4d$^2$, 3d$^7$4f$^2$, 3d$^7$4s5s, 3d$^7$4s4d, and 3d$^7$4s5d
for the even parity and 3d$^8$4p, 3d$^8$5p, 3d$^8$4f, 3d$^8$5f, 3d$^7$4s4p, 3d$^7$4s5p, 3d$^7$4s4f, 3d$^7$4s5f, and 3d$^7$4p4d 
for the odd parity. In this case, the semi-empirical process was performed to optimize the radial
integrals corresponding to 3d$^9$, 3d$^8$4s, and 3d$^8$4p configurations using the experimental
levels reported by \citet{sugarmusgrove1993}. The 3d$^9$4f levels were excluded from
the fit because many of these were found to be mixed with experimentally unknown levels
belonging notably to the 3d$^9$5p configuration.

The experimental and calculated energy levels for \Ion{Ge}{5}, expressed in cm$^{-1}$,  are reported in Table\,\ref{tab:gev:energy}
which also shows the differences between both sets of results ($\Delta E$) and, in the last column, the percentage 
composition in LS-coupling (only the first three components over 5\% are given). This last piece of information is useful 
because oscillator strengths for transitions connecting strongly perturbed levels are more sensitive to configuration interaction effects.

The calculated HFR oscillator strengths on a logarithmic scale ($\log gf$) and transition probabilities 
($gA$, in sec$^{-1}$) for \Ion{Ge} {5} are reported in Table\,\ref{tab:gev:loggf} with the corresponding wavelengths (in \AA) and 
energy levels (in cm$^{-1}$). In the last column, we give the cancellation factor CF as defined by \citet{cowan1981}. 
Low values of this factor indicate strong cancellation effects in the calculations. The corresponding transition 
probabilities could be very inaccurate so need to be considered with some care. It does appear, however, from the 
last column of the table that very few transitions are affected by such effects.

The experimental and calculated energy levels for \Ion{Ge}{6} appear in Table\,\ref{tab:gevi:energy} and the corresponding calculated 
HFR oscillator strengths and transition probabilities are reported in Table\,\ref{tab:gevi:loggf}. Here too, very few transitions are 
affected by cancellation effects so that for most of the transitions, the $f$ values should be reliable.

% Ge V  energy levels
\addtocounter{table}{1}

% Ge V  log gf
\addtocounter{table}{1}

% Ge VI energy levels
\addtocounter{table}{1}

% Ge VI log gf
\addtocounter{table}{1}

\section{Observations}
\label{sect:observation}

For our analysis, we mainly use the \emph{FUSE} spectrum of \re\ that
is described in detail by \citet{werneretal2012}.
In addition, we use UV spectra that were obtained with
ORFEUS\footnote{Orbiting Retrievable Far and Extreme Ultraviolet Spectrometer}/
\emph{BEFS}\footnote{Berkeley Extreme and Far-UV Spectrometer}, 
ORFEUS/
\emph{GHRS}\footnote{Goddard High-Resolution Spectrograph }
and
\emph{IUE}\footnote{International Ultraviolet Explorer}.
The \emph{BEFS} spectrum (909 - 1222\,\AA) 
is co-added from five observations
(ObsIds: BEFS2003, BEFS2126, BEFS2128, BEFS2133, BEFS2173; with a total observation time of 6826\,sec).
The \emph{GHRS} spectrum (1228 - 1275\,\AA, 1339 - 1375\,\AA, 1610 - 1655\,\AA)
is co-added from eight observations
(ObsIds: Z3GM0204T, Z3GM0205T, Z3JU0104T, Z3JU0107T, Z3JU0108T, Z3JU0109T, Z3JU010AT, Z3JU010BT; 5155\,sec).
The \emph{IUE} spectrum (1153 - 1947\,\AA) 
is the co-added spectrum
(ObsIds: SWP46428, SWP49788, SWP52796, SWP52803; 136\,193\,sec)
provided by
the \emph{IUE} NEWSIPS data base\footnote{\url{http://vega.lpl.arizona.edu/newsips/}} \citep{holbergetal1998}.

Optical spectra were taken in the framework of the
\emph{SPY}\footnote{ESO {\bf S}N\,Ia {\bf P}rogenitor surve{\bf Y}} project \citep{napiwotzkietal2001,napiwotzkietal2003} with 
\emph{UVES}\footnote{Ultraviolet and Visual Echelle Spectrograph} at 
ESO's\footnote{\url{http://www.eso.org/public/}} 
\emph{VLT}\footnote{Very Large Telescope}.

\section{The photospheric Ge abundance in \re}
\label{sect:model}

\begin{figure}[ht!]
   \resizebox{\hsize}{!}{\includegraphics{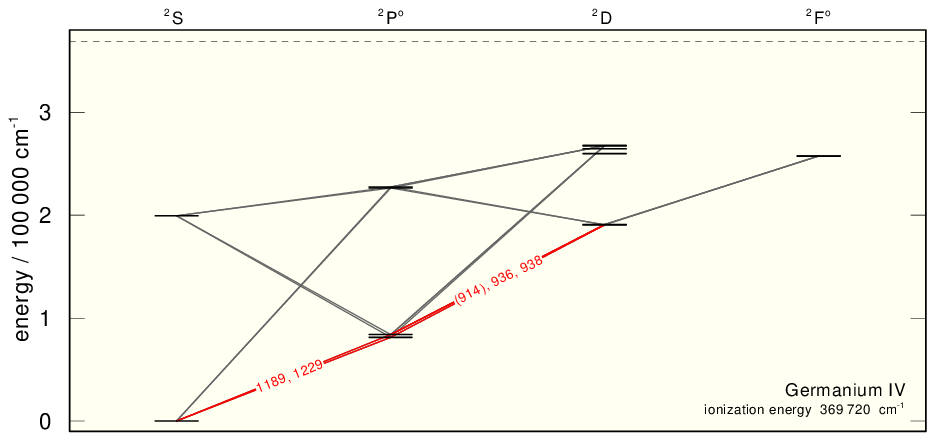}}
   \resizebox{\hsize}{!}{\includegraphics{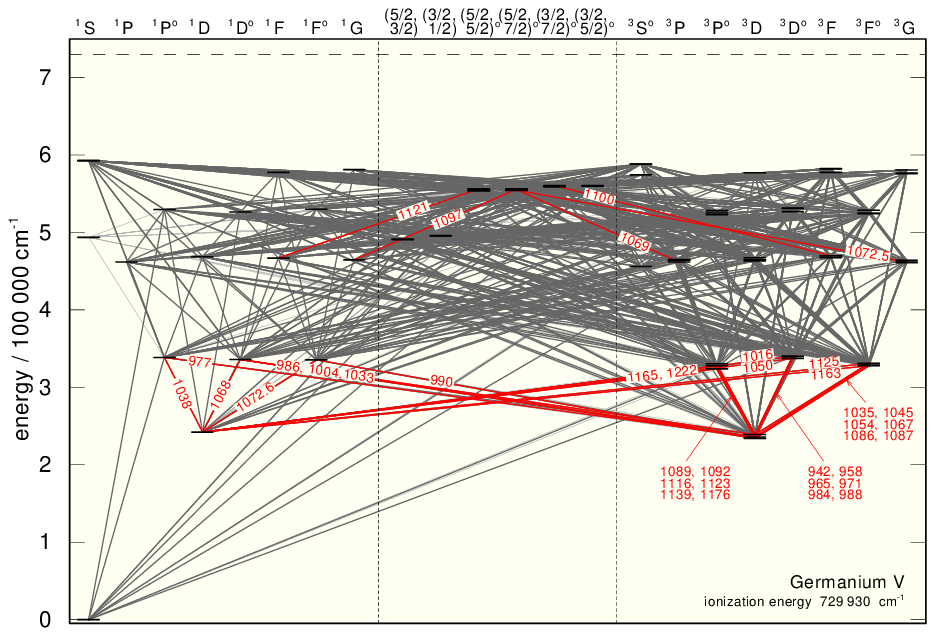}}
   \resizebox{\hsize}{!}{\includegraphics{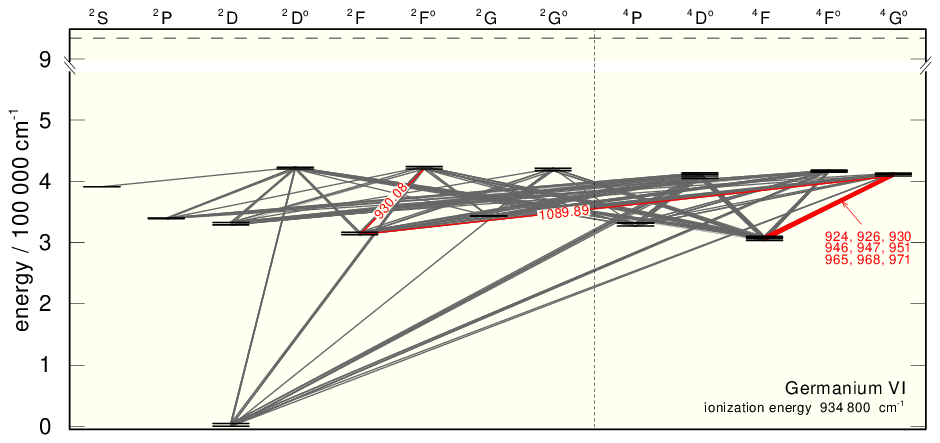}}
    \caption{Grotrian diagram of our 
             \Ion{Ge}{4} (top),
             \Ion{Ge}{5} (middle), and 
             \Ion{Ge}{6} (bottom) model ions.
             Thick black and thin gray lines represent radiative
             transitions with known and unknown $f$ values, respectively.
             The identified lines (red) are labeled with their respective wavelengths in \AA.
            }
   \label{fig:grotrian}
\end{figure}

\begin{figure*}[ht!]
   \resizebox{\hsize}{!}{\includegraphics{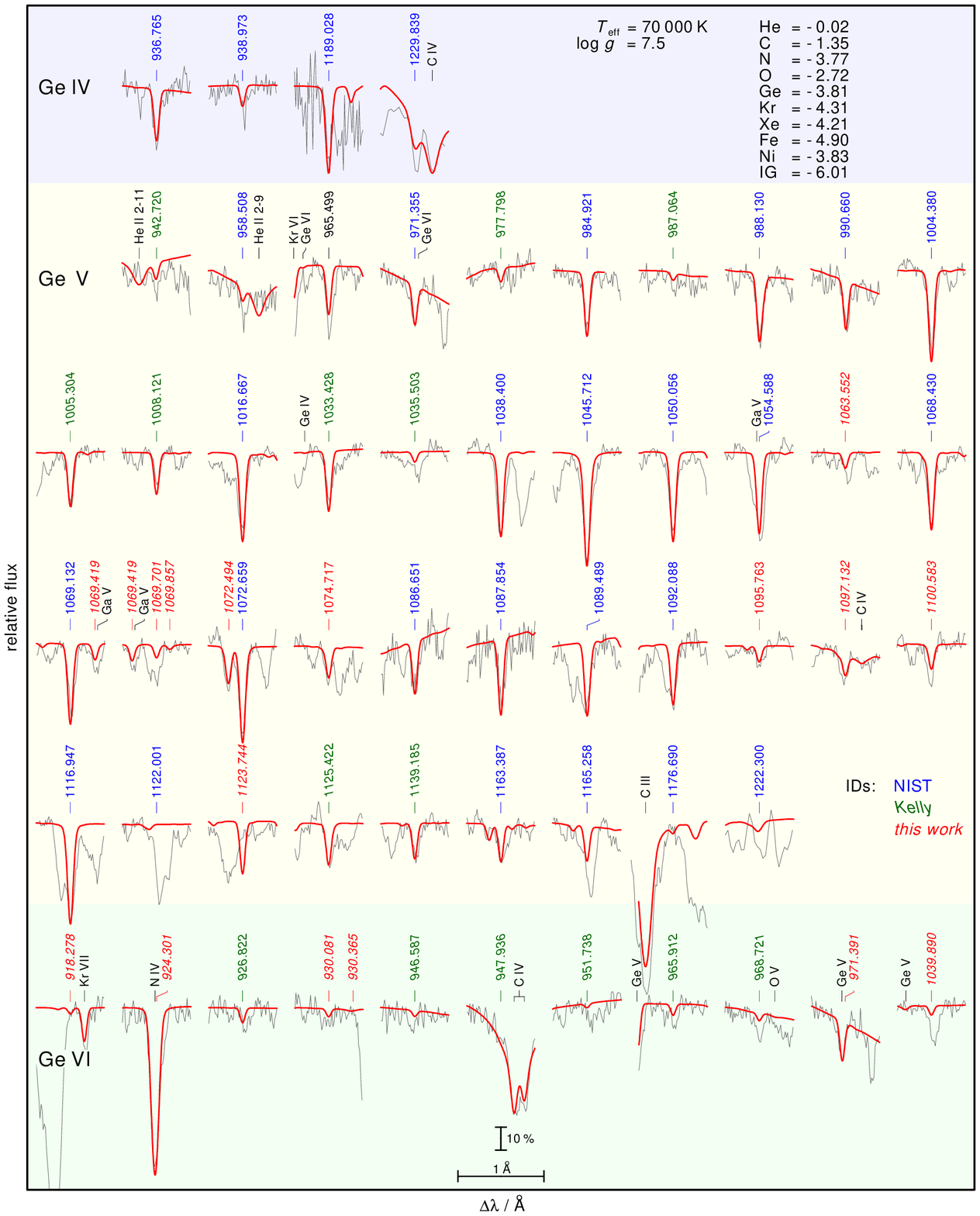}}
    \caption{\Ion{Ge}{4} (top), 
             \Ion{Ge}{5}, and 
             \Ion{Ge}{6} (bottom) lines
             in
             \emph{FUSE},
             ORFEUS\emph{/BEFS} (\Ionw{Ge}{4}{1189}), and
             \emph{IUE} (\Ionw{Ge}{5}{1222}, \Ionw{Ge}{4}{1229})
             observations compared with a \Teffw{70} / \loggw{7.5} \emph{TMAP} model.
             The abundances (top right) are logarithmic mass fractions.
             IG denotes a generic model atom \citep{rauchdeetjen2003}, which comprises
             Ca, Sc, Ti, V, Cr, Mn, and Co.
             The synthetic spectra a convolved with a Gaussian of
             0.05\,\AA\ (FWHM, 0.1\,\AA\ for the IUE comparison) to match the instrument resolution.
             A radial-velocity shift of $v_\mathrm{rad} = 23\,\mathrm{km/sec}$ is
             applied to the observation. 
            }
   \label{fig:ge70}
\end{figure*}

\Ion{Ge}{5} and \Ion{Ge}{6} are the dominant ionization stages in the line-forming region of
\re\ \sA{fig:ionization}. Thus, we constructed a Ge\,{\sc iii} - {\sc vii} model atom
(Table\,\ref{tab:ge}, Fig.~\ref{fig:grotrian}).
We used level energies from NIST for all ions.
For \Ion{Ge}{4}, we considered the oscillator strengths of \citet{nathduttamajumder2011}
and, where missing, approximated values from the isoelectronic \Ion{C}{4}.
\Ion{Ge}{5} and \Ion{Ge}{6} include our newly calculated oscillator strengths \sK{sect:getrans}.
Analogously to \citet{werneretal2012} in the case of Kr and Xe, the unknown $f$ values (Table\,\ref{tab:ge})
of these two ions were set to $10^{-4}$ within a spin and to $10^{-6}$ otherwise. 
Test calculations have shown that the Ge line profiles in the UV do not change when we set $f=0$ for these lines.
Photoionization rates were computed with hydrogen-like crosssections. Electron
collisional excitation and ionization rates were evaluated with the usual
approximation formulae following \citet{vanregemorter1962} and \citet{seaton1962},
respectively. 
This enabled us to build on the HeCNOKrXe models for \re\
described by \citet{werneretal2012} and to consider Ge opacities
as well as iron-group opacities (elements Ca - Ni, own determination of upper abundance limits) in addition.
Compared to \citet{werneretal2012},
we reduced the N abundance to match the \Ion{N}{4} 2p $^3$P$^\mathrm{o}$ - 2p$^2$ $^3$P  multiplet (921 - 924\,\AA,
Fig.\,\ref{fig:ge70}).

\begin{table}[ht!]
\caption{Statistics of the Ge model atom used in our calculations.}
\label{tab:ge}
\begin{tabular}{lrrcrrr}
\hline
\hline
\noalign{\smallskip}
           & \multicolumn{2}{c}{levels} &~& \multicolumn{3}{c}{line transitions}        \\
\cline{2-3}
\cline{5-7}
                     &      &     & &       &           &              \vspace{-2.0em}\\
\noalign{\smallskip}
                 ion &      &     & &       &           &               \vspace{-0.3em}\\
                     & NLTE & LTE & & total & known $f$ & unknown $f$                  \\
\hline
\noalign{\smallskip}
\Ion{}{3}            &   14\hspace{2mm}\hbox{} &   2\hspace{2mm}\hbox{} & &     0 &                        &                        \\
\Ion{}{4}            &    8\hspace{2mm}\hbox{} &   1\hspace{2mm}\hbox{} & &     8 &   8\hspace{3mm}\hbox{} &                        \\
\Ion{}{5}            &   85\hspace{2mm}\hbox{} &   0\hspace{2mm}\hbox{} & &  1345 & 878\hspace{3mm}\hbox{} & 467\hspace{5mm}\hbox{} \\
\Ion{}{6}            &   36\hspace{2mm}\hbox{} &   0\hspace{2mm}\hbox{} & &   235 & 160\hspace{3mm}\hbox{} &  75\hspace{5mm}\hbox{} \\
\Ion{}{7}            &    1\hspace{2mm}\hbox{} &   0\hspace{2mm}\hbox{} & &     0 &                        &                        \\
\hline
\end{tabular}
\end{table}

We used the \emph{T\"ubingen Model-Atmosphere Package}
\citep[\emph{TMAP}\footnote{\url{http://astro.uni-tuebingen.de/~TMAP}}][]{werneretal2003}
to calculate state-of-the-art, plane-parallel, chemically homogeneous model atmospheres in
hydrostatic and radiative equilibrium. The considered model atoms are those that are provided via
the \emph{T\"ubingen Model-Atom Database}
\citep[\emph{TMAD}\footnote{\url{http://astro.uni-tuebingen.de/~TMAD}},][]{rauchdeetjen2003}.

We compared the available UV spectra of \re\ \sK{sect:observation}
with our \emph{TMAP} model
(and wavelength positions given by NIST and Kelly's database)
in order to identify Ge lines.
The line lists in that energy region,
especially for the highly ionized trans-iron elements that are encountered,
are rather incomplete, and thus, lines that are not considered in the
models may contribute to the assumed isolated Ge lines.
\Ionww{Ga}{5}{1054.560, 1069.450} are two examples (Fig.~\ref{fig:ge70}).
The consideration of Ga in the model-atmosphere calculation would improve
the fit of the blends with \Ionww{Ge}{5}{1054.588, 1069.419}.
However, we identify four \Ion{Ge}{4}, 37 \Ion{Ge}{5}, and six \Ion{Ge}{6}
lines (Table\,\ref{tab:gelines}).
All \Ion{Ge}{4} and \Ion{Ge}{5} lines are in general reproduced in both 
strength and width by our model 
simultaneously at $\log \mathrm{Ge} = -3.81 \pm 0.3$ (Fig.~\ref{fig:ge70}). 
There is only one line, \Ionw{Ge}{5}{1123.744} that is much too strong in
our model. The reason is unknown.
The \Ion{Ge}{6} lines are weak in our model and just emerge from  the noise in the
\emph{FUSE} observation, but they do agree. \Ionww{Ge}{6}{986.721, 1039.890}
are too weak to reproduce the observed absorption features, most likely due
to unknown blends at their positions that are not considered in the model.
However, the large number of identified Ge lines and their modeling give convincing
evidence that our model and the determined Ge abundance are realistic.

\begin{table*}[ht!]
\caption{Identified Ge lines in the UV spectrum of \re. }
\label{tab:gelines}
\begin{tabular}{lccrrr@{.}ll}
\hline
\hline
\noalign{\smallskip}
ion               & lower level          & upper level            & & \multicolumn{1}{c}{$f$ value} & \multicolumn{2}{c}{wavelength / \AA} & comment \\
\noalign{\smallskip}
\hline
\noalign{\smallskip}
Ge\,{\sc iv}      & 4p $^2$P$^\mathrm{o}_{3/2}$ & 4d $^2$D$^\mathrm{ }_{5/2}$   & &  8.91E-01 & \hbox{}\hspace{5mm}  936&765  &                    \\
                  & 4p $^2$P$^\mathrm{o}_{3/2}$ & 4d $^2$D$^\mathrm{ }_{3/2}$   & &  9.89E-02 &                      938&973  &                    \\
                  & 4s $^2$S$^\mathrm{ }_{1/2}$ & 4p $^2$P$^\mathrm{o}_{3/2}$   & &  5.54E-01 &                     1189&028  &                    \\
                  & 4s $^2$S$^\mathrm{ }_{1/2}$ & 4p $^2$P$^\mathrm{o}_{1/2}$   & &  2.66E-01 &                     1229&839  &                    \\
\noalign{\smallskip}                                                                                                                                 
\hline                                                                                                                                               
\noalign{\smallskip}                                                                                                                                 
Ge\,{\sc v }      & 4s $^3$D$^\mathrm{ }_{3}$   & 4p $^3$D$^\mathrm{o}_{2}$     & &  8.22E-03 &                      942&720  &                    \\
                  & 4s $^3$D$^\mathrm{ }_{2}$   & 4p $^3$D$^\mathrm{o}_{2}$     & &  3.17E-02 &                      958&508  &                    \\
                  & 4s $^3$D$^\mathrm{ }_{2}$   & 4p $^3$D$^\mathrm{o}_{1}$     & &  3.56E-02 &                      965&499  &                    \\
                  & 4s $^3$D$^\mathrm{ }_{3}$   & 4p $^3$D$^\mathrm{o}_{3}$     & &  1.22E-01 &                      971&355  &                    \\
                  & 4s $^3$D$^\mathrm{ }_{2}$   & 4p $^1$P$^\mathrm{o}_{1}$     & &  3.56E-03 &                      977&798  & Kelly wavelength   \\
                  & 4s $^3$D$^\mathrm{ }_{1}$   & 4p $^3$D$^\mathrm{o}_{2}$     & &  1.08E-01 &                      984&921  &                    \\
                  & 4s $^3$D$^\mathrm{ }_{3}$   & 4p $^1$D$^\mathrm{o}_{2}$     & &  1.13E-03 &                      987&064  & Kelly wavelength   \\
                  & 4s $^3$D$^\mathrm{ }_{2}$   & 4p $^3$D$^\mathrm{o}_{3}$     & &  1.00E-01 &                      988&130  &                    \\
                  & 4s $^3$D$^\mathrm{ }_{3}$   & 4p $^1$F$^\mathrm{o}_{3}$     & &  1.13E-01 &                      990&660  &                    \\
                  & 4s $^3$D$^\mathrm{ }_{2}$   & 4p $^1$D$^\mathrm{o}_{2}$     & &  1.70E-01 &                     1004&380  & NIST wavelength    \\
                  & 4s $^3$D$^\mathrm{ }_{1}$   & 4p $^1$P$^\mathrm{o}_{1}$     & &  5.41E-02 &                     1005&304  & Kelly wavelength   \\
                  & 4s $^3$D$^\mathrm{ }_{2}$   & 4p $^1$F$^\mathrm{o}_{3}$     & &  1.21E-02 &                     1008&121  &                    \\
                  & 4s $^1$D$^\mathrm{ }_{2}$   & 4p $^3$D$^\mathrm{o}_{2}$     & &  1.95E-01 &                     1016&667  &                    \\
                  & 4s $^3$D$^\mathrm{ }_{1}$   & 4p $^1$D$^\mathrm{o}_{2}$     & &  6.96E-02 &                     1033&428  & Kelly wavelength   \\
                  & 4s $^3$D$^\mathrm{ }_{3}$   & 4p $^3$F$^\mathrm{o}_{2}$     & &  1.01E-03 &                     1035&503  &                    \\
                  & 4s $^1$D$^\mathrm{ }_{2}$   & 4p $^1$P$^\mathrm{o}_{1}$     & &  1.45E-01 &                     1038&400  & NIST wavelength    \\
                  & 4s $^3$D$^\mathrm{ }_{3}$   & 4p $^3$F$^\mathrm{o}_{4}$     & &  3.94E-01 &                     1045&712  &                    \\
                  & 4s $^1$D$^\mathrm{ }_{2}$   & 4p $^3$D$^\mathrm{o}_{3}$     & &  1.56E-01 &                     1050&056  &                    \\
                  & 4s $^3$D$^\mathrm{ }_{2}$   & 4p $^3$F$^\mathrm{o}_{2}$     & &  8.93E-02 &                     1054&588  &                    \\
                  & 4d $^3$P$^\mathrm{ }_{1}$   & 4f $^5$F$^\mathrm{o}_{2}$     & &  8.57E-01 &                     1063&552  &                    \\
                  & 4s $^1$D$^\mathrm{ }_{2}$   & 4p $^1$D$^\mathrm{o}_{2}$     & &  8.73E-02 &                     1068&430  & NIST wavelength    \\
                  & 4s $^3$D$^\mathrm{ }_{3}$   & 4p $^3$F$^\mathrm{o}_{3}$     & &  6.53E-02 &                     1069&132  & NIST wavelength    \\
                  & 4d $^3$P$^\mathrm{ }_{2}$   & 4f                            & &  5.02E-01 &                     1069&419  &                    \\
                  & 4d $^1$P$^\mathrm{ }_{1}$   & 4f                            & &  6.07E-01 &                     1069&701  &                    \\
                  & 4d $^3$P$^\mathrm{ }_{2}$   & 4f                            & &  1.10E-01 &                     1069&857  &                    \\
                  & 4d $^3$G$^\mathrm{ }_{5}$   & 4f                            & &  9.97E-01 &                     1072&494  &                    \\
                  & 4s $^1$D$^\mathrm{ }_{2}$   & 4p $^1$F$^\mathrm{o}_{3}$     & &  2.52E-01 &                     1072&659  &                    \\
                  & 4d $^3$G$^\mathrm{ }_{4}$   & 4f                            & &  9.68E-01 &                     1074&717  &                    \\
                  & 4s $^3$D$^\mathrm{ }_{1}$   & 4p $^3$F$^\mathrm{o}_{2}$     & &  3.18E-01 &                     1086&651  &                    \\
                  & 4s $^3$D$^\mathrm{ }_{2}$   & 4p $^3$F$^\mathrm{o}_{3}$     & &  3.03E-01 &                     1087&854  &                    \\
                  & 4s $^3$D$^\mathrm{ }_{2}$   & 4p $^3$P$^\mathrm{o}_{1}$     & &  1.42E-01 &                     1089&489  &                    \\
                  & 4s $^3$D$^\mathrm{ }_{1}$   & 4p $^3$P$^\mathrm{o}_{0}$     & &  1.01E-01 &                     1092&088  &                    \\
                  & 4d $^3$G$^\mathrm{ }_{3}$   & 4f $^5$D$^\mathrm{o}_{3}$     & &  1.53E-02 &                     1095&763  &                    \\
                  & 4d $^1$G$^\mathrm{ }_{4}$   & 4f                            & &  7.69E-01 &                     1097&132  &                    \\
                  & 4d $^3$F$^\mathrm{ }_{3}$   & 4f                            & &  9.01E-01 &                     1100&583  &                    \\
                  & 4s $^3$D$^\mathrm{ }_{3}$   & 4p $^3$P$^\mathrm{o}_{2}$     & &  1.97E-01 &                     1116&947  &                    \\
                  & 4d $^1$F$^\mathrm{ }_{3}$   & 4f                            & &  9.44E-02 &                     1122&001  & NIST wavelength, very weak  \\
                  & 4s $^3$D$^\mathrm{ }_{1}$   & 4p $^3$P$^\mathrm{o}_{1}$     & &  3.83E-02 &                     1123&744  &                    \\
                  & 4s $^1$D$^\mathrm{ }_{2}$   & 4p $^3$F$^\mathrm{o}_{2}$     & &  1.29E-02 &                     1125&422  &                    \\
                  & 4s $^3$D$^\mathrm{ }_{2}$   & 4p $^3$P$^\mathrm{o}_{2}$     & &  9.14E-03 &                     1139&185  &                    \\
                  & 4s $^1$D$^\mathrm{ }_{2}$   & 4p $^3$F$^\mathrm{o}_{3}$     & &  1.29E-02 &                     1163&387  &                    \\
                  & 4s $^1$D$^\mathrm{ }_{2}$   & 4p $^3$P$^\mathrm{o}_{1}$     & &  1.48E-02 &                     1165&258  &                    \\
                  & 4s $^3$D$^\mathrm{ }_{1}$   & 4p $^3$P$^\mathrm{o}_{2}$     & &  2.06E-03 &                     1176&690  & \Ion{C}{3} blend   \\
                  & 4s $^1$D$^\mathrm{ }_{2}$   & 4p $^3$P$^\mathrm{o}_{2}$     & &  4.91E-03 &                     1222&300  &                    \\
\noalign{\smallskip}                                                                                                                                 
\hline                                                                                                                                               
\noalign{\smallskip}                                                                                                                                 
Ge\,{\sc vi}      & 4s $^4$F$^\mathrm{ }_{7/2}$ & 4p $^4$G$^\mathrm{o}_{\,9/2}$ & &  2.34E-04 &                      918&278  &                    \\
                  & 4s $^4$F$^\mathrm{ }_{9/2}$ & 4p $^4$G$^\mathrm{o}_{\,7/2}$ & &  2.34E-04 &                      924&301  & \Ion{N}{4} blend   \\
                  & 4s $^4$F$^\mathrm{ }_{9/2}$ & 4p $^4$G$^\mathrm{o}_{ 11/2}$ & &  3.63E-01 &                      926&822  &                    \\
                  & 4s $^2$F$^\mathrm{ }_{7/2}$ & 4p $^2$F$^\mathrm{o}_{\,7/2}$ & &  2.43E-01 &                      930&081  &                    \\
                  & 4s $^4$F$^\mathrm{ }_{7/2}$ & 4p $^4$G$^\mathrm{o}_{\,5/2}$ & &  8.07E-04 &                      930&365  &                    \\
                  & 4s $^4$F$^\mathrm{ }_{7/2}$ & 4p $^4$G$^\mathrm{o}_{\,7/2}$ & &  1.28E-01 &                      946&587  &                    \\
                  & 4s $^4$F$^\mathrm{ }_{9/2}$ & 4p $^4$G$^\mathrm{o}_{\,9/2}$ & &  1.15E-01 &                      947&936  & \Ion{C}{4} blend   \\
                  & 4s $^4$F$^\mathrm{ }_{5/2}$ & 4p $^4$G$^\mathrm{o}_{\,5/2}$ & &  1.03E-01 &                      951&738  &                    \\
                  & 4s $^4$F$^\mathrm{ }_{3/2}$ & 4p $^4$G$^\mathrm{o}_{\,5/2}$ & &  2.62E-01 &                      965&912  &                    \\
                  & 4s $^4$F$^\mathrm{ }_{5/2}$ & 4p $^4$G$^\mathrm{o}_{\,7/2}$ & &  1.96E-01 &                      968&721  &                    \\
                  & 4s $^4$F$^\mathrm{ }_{7/2}$ & 4p $^4$G$^\mathrm{o}_{\,9/2}$ & &  1.81E-01 &                      971&391  &                    \\
                  & 4s $^2$F$^\mathrm{ }_{7/2}$ & 4p $^4$G$^\mathrm{o}_{\,9/2}$ & &  1.81E-01 &                     1039&890  &                    \\
\noalign{\smallskip}
\hline
\end{tabular}

\noindent
{\footnotesize 
      Note: The $f$ values of \Ion{Ge}{4} are from \citet{nathduttamajumder2011}.
      All Ge\,{\sc v} and {\sc vi} wavelengths are calculated from energy levels in
      Tables \ref{tab:gev:energy} and \ref{tab:gevi:energy} unless otherwise mentioned in the comment column.}
\end{table*}

\section{Effective temperature and surface gravity of \re}
\label{sect:tefflogg}

Both \Teff\ and \logg\ were adopted from \citet{werneretal2012} for this analysis.
Figure~\ref{fig:ge70} shows that the \Ion{C}{3} multiplet 2p $^3$P$^\mathrm{o}$ - 2p$^2$ $^3$P
(1174 - 1176\,\AA) in our model is too weak. Since an increased C abundance
would strengthen \Ion{C}{4} lines as well, e.g\@. \Ionww{C}{4}{948.09, 948.21} (Fig.~\ref{fig:ge70}),
this is evidence that \Teff\ of the model is too high and/or \logg\ is too low. A respective variation
would change the \Ion{C}{3} / \Ion{C}{4} ionization equilibrium towards the lower ionization and
improve the agreement of the \Ion{C}{3} lines.
Figures \ref{fig:teff70} and \ref{fig:teff65} demonstrate this for \Teffw{70} and \Teffw{65}.
We compared theoretical
\Ion{He}{1}, \Ion{He}{2}, \Ion{C}{3}, \Ion{C}{4}, \Ion{O}{4}, and \Ion{O}{5} line profiles
with \emph{FUSE} and \emph{GHRS} UV observations and optical \emph{UVES} observations.
The most prominent \Ionw{C}{3}{977.020} has a strong, blue-shifted interstellar component and is 
not well suited to an analysis. 
The better agreement of \Ionww{C}{3}{1175} in the line cores favors \Teffw{65},
while the ``shoulders'' between the \Ionww{C}{3}{1175} components are better matched at \Teffw{70}.
The lower \Teff\ is supported by the
\Ion{Ge}{4} / \Ion{Ge}{5} ionization balance (Fig.\,\ref{fig:ge65}), if we judge e.g\@.
\Ionw{Ge}{4}{936.765}. The \Ion{Ge}{5} lines appear almost with same strengths in both
(\Teffw{70} and \Teffw{65}) models. On the other hand, \Ion{Kr}{6} / \Ion{Kr}{7} favors
\Teffw{70} \citep{werneretal2012}, and \Ionw{He}{1}{4471} is too strong in the model at
\Teffw{65} (Fig.\,\ref{fig:ge65}). The \Ion{O}{4} / \Ion{O}{5} ionization appears unchanged
between \Teffw{65} and \Teffw{70}.

% Teff 70kK
\addtocounter{figure}{1}

% Teff 65kK
\addtocounter{figure}{1}

% ge   65kK
\addtocounter{figure}{1}

It is worthwhile mentioning that a lower \Teff\ would strongly improve the
agreement between model and the observed EUV flux \citep[\object{EUVE\,J0503$-$28.8},][]{werneretal2001}.
A more precise determination of \Teff\ and \logg\ of \re\ 
based on additional high-S/N optical spectra and more
ionization equilibria of different species is highly desirable.
Our test calculations have shown that our Ge line identifications 
and abundance determination are affected only marginally by 
this uncertainty in atmosphere parameters.

\section{Results and conclusions}
\label{sect:results}

Successful reproduction of the identified Ge lines in
high-resolution UV spectra of \re\ by our
synthetic spectra calculated from NLTE model atmospheres
using newly calculated oscillator strengths of \Ion{Ge}{5} and \Ion{Ge}{6}
shows that ---  when done with sufficient care --- theory works.

We derive a photospheric abundance of $\log \mathrm{Ge} = -3.8 \pm 0.3$ (mass fraction)
in \re. This is about 650 times the solar abundance. This high value is similar to
the results of \citet{werneretal2012} for
Kr (450 times solar) and
Xe (3800 times solar).

The identifications of trans-iron elements in the \emph{FUSE} spectrum of \re\
and the abundance determinations of Ge, Kr, and Xe show that \re\ is important
for our understanding of the non-DA
white dwarf evolutionary channel. Further abundance determinations of the
identified species is highly desirable. This is a challenge for atomic physicists.

It is worthwhile mentioning here the two HST observations of \re\ taken with 
\emph{STIS}\footnote{Space Telescope Imaging Spectrograph}
(1999-03-23, ObsIds O56401010, O56401020) that missed the star because the
prior target acquisition apparently failed. Unfortunately, they were not repeated.
The available \emph{GHRS}\footnote{Goddard High Resolution Spectrograph} 
observations cover only small wavelength sections of
the NUV, and the \emph{IUE} high-resolution spectra (e.g\@. SWP52803HL) have too-low an
S/N. Obtaining high-resolution, high S/N spectra with
HST/\emph{STIS} should not be missed because the NUV spectrum probably offers important, additional
spectral information.

The Ge model ions that were used in this analysis were developed in the framework 
of the \emph{Virtual Observatory} (\emph{VO}\footnote{\url{http://www.ivoa.net/}})
in a \emph{German Astrophysical Virtual Observatory}
(\emph{GAVO}\footnote{\url{http://www.g-vo.org}}) project and are provided
within \emph{TMAD} (Sect.\,\ref{sect:model}). 
The spectral energy distribution of our final model
can be retrieved in \emph{VO}-compliant form via the
registered \emph{VO} service \emph{TheoSSA}\footnote{\url{http://dc.g-vo.org/theossa}}.

\begin{acknowledgements}
TR is supported by the German Aerospace Center (DLR, grant 05\,OR\,0806).
Financial support from the Belgian FRS-FNRS is also acknowledged. 
EB and PQ are Research Director and Senior Research Associate, respectively, of this organization.
This research has made use of the SIMBAD database, operated at the CDS, Strasbourg, France.
We thank Ralf Napiwotzki for providing us the \emph{SPY} spectrum of \re.
Some of the data presented in this paper were obtained from the
Mikulski Archive for Space Telescopes (MAST). STScI is operated by the
Association of Universities for Research in Astronomy, Inc., under NASA
contract NAS5-26555. Support for MAST for non-HST data is provided by
the NASA Office of Space Science via grant NNX09AF08G and by other
grants and contracts. 
\end{acknowledgements}

\bibliographystyle{aa}
\bibliography{20014}

\onlfig{4}{
\begin{figure*}[ht!]
   \resizebox{\hsize}{!}{\includegraphics{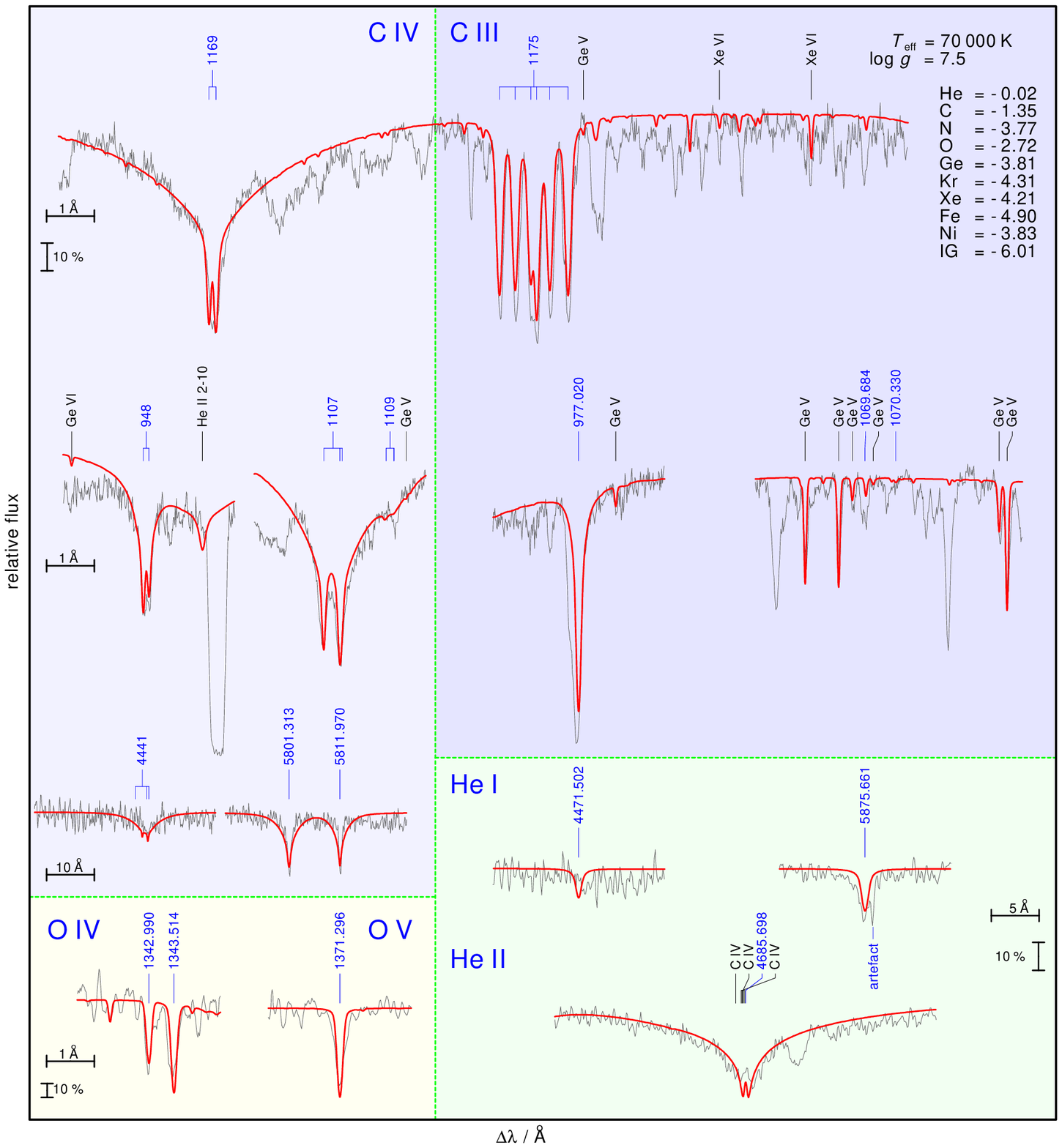}}
    \caption{\Ion{He}{1} / \Ion{He}{2},
             \Ion{C}{3}  / \Ion{C}{4}, and
             \Ion{O}{4}  / \Ion{O}{5} 
             ionization equilibria in our \Teffw{70} model compared
             with \emph{FUSE}, 
             \emph{GHRS} \citep[smoothed with a low-pass filter ($m=15$, $n=4$) for clarity,][]{savitzkygolay1964}, and
             \emph{UVES} observations.
             Wavelength and flux scales are indicated by bars.
             The model spectrum is convolved with a Gaussian to match the respective instrument's resolution.
             Unidentified lines in the model stem from Ca - Ni.}
   \label{fig:teff70}
\end{figure*}
}

\onlfig{5}{
\begin{figure*}[ht!]
   \resizebox{\hsize}{!}{\includegraphics{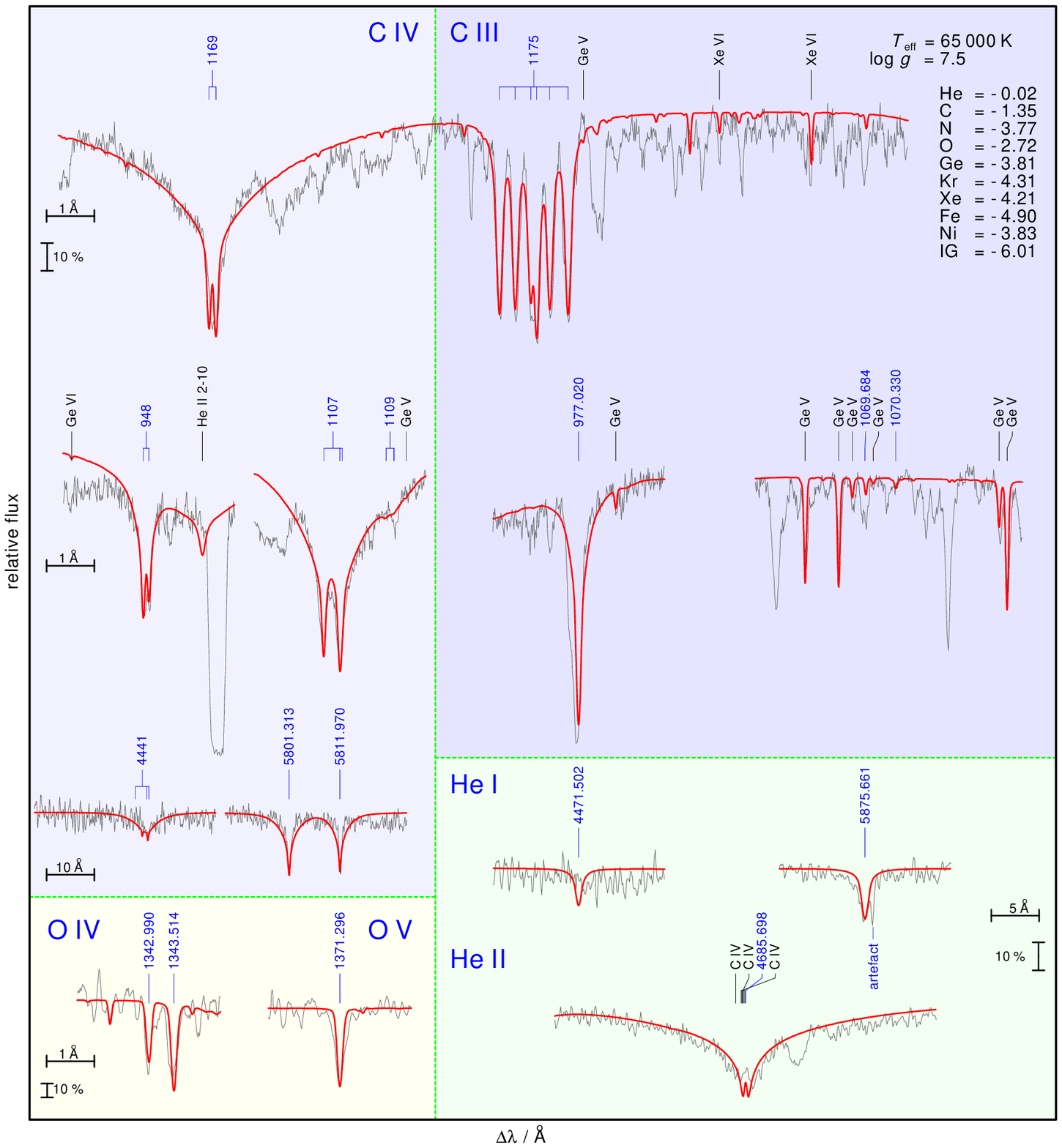}}
    \caption{Same as Fig.\,\ref{fig:teff70} for \Teffw{65}.
            }
   \label{fig:teff65}
\end{figure*}
}

\onlfig{6}{
\begin{figure*}[ht!]
   \resizebox{\hsize}{!}{\includegraphics{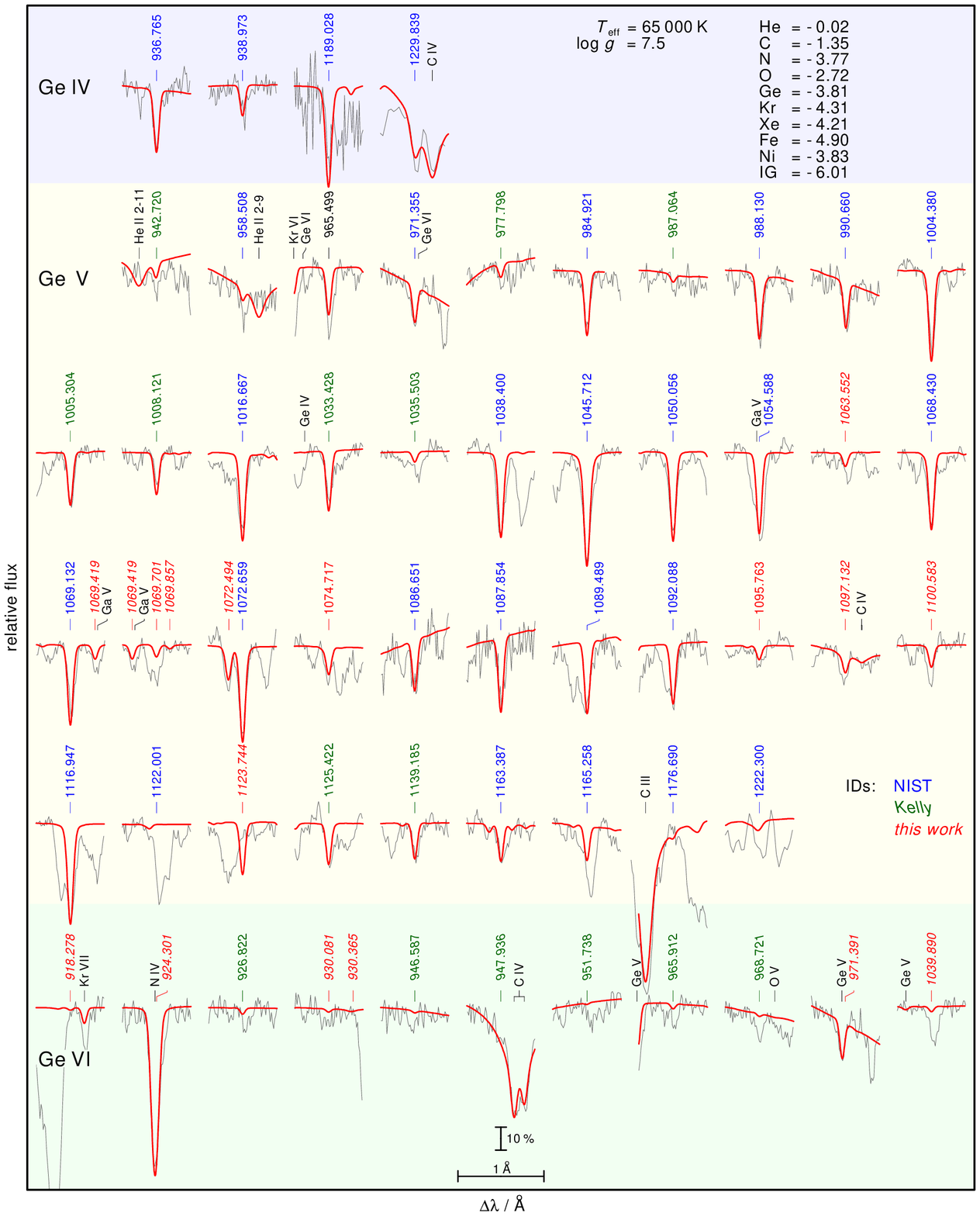}}
    \caption{Same as Fig.\,\ref{fig:ge70}  for \Teffw{65}.
            }
   \label{fig:ge65}
\end{figure*}
}

\onltab{1}{
\longtab{1}{
\begin{longtable}{rrrcl}
\caption{\label{tab:gev:energy}Energy levels of \Ion{Ge}{5} (in cm$^{-1}$). The first three LS-components are given when they are over 5\,\%.}\\
\hline\hline
\noalign{\smallskip}
E$_{exp}$ & E$_{calc}$ & $\Delta$$E$ & $J$ & LS-coupling composition (\%) \\
\noalign{\smallskip}
\hline
\endfirsthead
\caption{continued.}\\
\hline\hline
\noalign{\smallskip}
E$_{exp}$ & E$_{calc}$ & $\Delta$$E$ & $J$ & LS-coupling composition (\%) \\
\noalign{\smallskip}
\hline
\noalign{\smallskip}
\endhead
\hline
\noalign{\smallskip}
\endfoot
\noalign{\smallskip}
              0.0 &      0 &    0 &  0 &  99 3d$^{10}$ $^1$S  \\
         234219.3 & 234226 &   -7 &  3 &  99 3d$^9$4s $^3$D  \\
         235967.0 & 235958 &    9 &  2 &  83 3d$^9$4s $^3$D  \\
         238764.9 & 238765 &   -4 &  1 &  99 3d$^9$4s $^3$D  \\
         241935.2 & 241937 &   -2 &  2 &  83 3d$^9$4s $^1$D  \\
         323748.9 & 323877 & -128 &  2 &  96 3d$^9$4p $^3$P$^o$  \\
         327753.0 & 327889 & -136 &  1 &  96 3d$^9$4p $^3$P$^o$  \\
         327891.0 & 327792 &   99 &  3 &  68 3d$^9$4p $^3$F$^o$ + 27 3d$^9$4p $^1$F$^o$  \\
\hline
\end{longtable}
}
}

\onltab{2}{
\longtab{2}{
\begin{longtable}{rrccrrccrrr}
\caption{\label{tab:gev:loggf} Calculated HFR oscillator strengths (log gf) and transition probabilities (gA, in sec${-1}$) in \Ion{Ge}{5}. CF is the cancellation factor as defined by \citet{cowan1981}. In columns 3 and 6, $e$ is written for even and $o$ for odd.}\\
\hline\hline
\noalign{\smallskip}
                 & \multicolumn{3}{c}{Lower level} && \multicolumn{3}{c}{Upper level} &           &      &    \\
\cline{2-4}\cline{6-8}
\noalign{\smallskip}
Wavelength / \AA & Energy / cm$^{-1}$ & Parity & $j$  && Energy / cm$^{-1}$ & Parity & $j$  & $\log gf$ & $gA$ & CF \\
\noalign{\smallskip}
\hline
\endfirsthead
\caption{continued.}\\
\hline\hline
\noalign{\smallskip}
                 & \multicolumn{3}{c}{Lower level} && \multicolumn{3}{c}{Upper level} &           &      &    \\
\cline{2-4}\cline{6-8}
\noalign{\smallskip}
Wavelength / \AA & Energy / cm$^{-1}$ & Parity & $j$  && Energy / cm$^{-1}$ & Parity & $j$  & $\log gf$ & $gA$ & CF \\
\noalign{\smallskip}
\hline
\noalign{\smallskip}
\endhead
\hline
\noalign{\smallskip}
\endfoot
\noalign{\smallskip}
     150.433 &        0 & e &  0 & &   664749 & o &  1 &   -2.88 &  3.85E+08 &   0.039 \\
     151.488 &        0 & e &  0 & &   660117 & o &  1 &   -2.82 &  4.42E+08 &  -0.054 \\
     152.175 &        0 & e &  0 & &   657140 & o &  1 &   -0.76 &  4.96E+10 &  -0.382 \\
     153.196 &        0 & e &  0 & &   652760 & o &  1 &   -0.82 &  4.28E+10 &   0.391 \\
     153.997 &        0 & e &  0 & &   649365 & o &  1 &   -2.37 &  1.19E+09 &  -0.041 \\
     155.406 &        0 & e &  0 & &   643477 & o &  1 &   -3.43 &  1.04E+08 &   0.067 \\
     157.935 &        0 & e &  0 & &   633171 & o &  1 &   -3.61 &  6.51E+07 &  -0.062 \\
     160.081 &        0 & e &  0 & &   624684 & o &  1 &   -3.30 &  1.30E+08 &   0.350 \\
\hline
\end{longtable}
}
}

\onltab{3}{
\longtab{3}{
\begin{longtable}{rrrcl}
\caption{\label{tab:gevi:energy}Energy levels of \Ion{Ge}{6} (in cm$^{-1}$). The first three LS-components are given when they are over 5\,\%.}\\
\hline\hline
\noalign{\smallskip}
E$_{exp}$ & E$_{calc}$ & $\Delta$$E$ & $J$ & LS-coupling composition (\%) \\
\noalign{\smallskip}
\hline
\endfirsthead
\caption{continued.}\\
\hline\hline
\noalign{\smallskip}
E$_{exp}$ & E$_{calc}$ & $\Delta$$E$ & $J$ & LS-coupling composition (\%) \\
\noalign{\smallskip}
\hline
\noalign{\smallskip}
\endhead
\hline
\noalign{\smallskip}
\endfoot
\noalign{\smallskip}
              0.0 &      0 &     0 &  2.5 &  99 3d$^9$ $^2$D  \\
           4560.0 &   4560 &     0 &  1.5 &  99 3d$^9$ $^2$D  \\
         303696.2 & 303702 &    -6 &  4.5 &  99 3d$^8$($^3$F)4s $^4$F \\
         306243.4 & 306227 &    17 &  3.5 &  93 3d$^8$($^3$F)4s $^4$F + 7 3d$^8$($^3$F)4s $^2$F \\
         308657.2 & 308619 &    38 &  2.5 &  97 3d$^8$($^3$F)4s $^4$F \\
         310199.1 & 310149 &    50 &  1.5 &  97 3d$^8$($^3$F)4s $^4$F \\
         313024.6 & 313083 &   -59 &  3.5 &  93 3d$^8$($^3$F)4s $^2$F + 7 3d$^8$($^3$F)4s $^4$F  \\
         316936.8 & 316978 &   -41 &  2.5 &  94 3d$^8$($^3$F)4s $^2$F \\
\hline
\end{longtable}
}
}

\onltab{4}{
\longtab{4}{
\begin{longtable}{rrccrrccrrr}
\caption{\label{tab:gevi:loggf}Calculated HFR oscillator strengths (log gf) and transition probabilities (gA, in sec${-1}$) in \Ion{Ge}{6}. CF is the cancellation factor as defined by \citet{cowan1981}. In columns 3 and 6, $e$ is written for even and $o$ for odd.}\\
\hline\hline
\noalign{\smallskip}
                 & \multicolumn{3}{c}{Lower level} && \multicolumn{3}{c}{Upper level} &           &      &    \\
\cline{2-4}\cline{6-8}
\noalign{\smallskip}
Wavelength / \AA & Energy / cm$^{-1}$ & Parity & $j$  && Energy / cm$^{-1}$ & Parity & $j$  & $\log gf$ & $gA$ & CF \\
\noalign{\smallskip}
\hline
\endfirsthead
\caption{continued.}\\
\hline\hline
\noalign{\smallskip}
                 & \multicolumn{3}{c}{Lower level} && \multicolumn{3}{c}{Upper level} &           &      &    \\
\cline{2-4}\cline{6-8}
\noalign{\smallskip}
Wavelength / \AA & Energy / cm$^{-1}$ & Parity & $j$  && Energy / cm$^{-1}$ & Parity & $j$  & $\log gf$ & $gA$ & CF \\
\noalign{\smallskip}
\hline
\noalign{\smallskip}
\endhead
\hline
\noalign{\smallskip}
\endfoot
\noalign{\smallskip}
     199.770 &        0 & e &  2.5 & & 500575 & o &  1.5 & -1.20 &  1.06E+10 &   0.408 \\
     201.607 &     4560 & e &  1.5 & & 500575 & o &  1.5 & -2.00 &  1.66E+09 &  -0.421 \\
     203.409 &     4560 & e &  1.5 & & 496180 & o &  0.5 & -1.30 &  8.01E+09 &   0.471 \\
     218.161 &        0 & e &  2.5 & & 458378 & o &  3.5 & -1.77 &  2.40E+09 &   0.466 \\
     221.483 &        0 & e &  2.5 & & 451501 & o &  1.5 & -3.80 &  2.15E+07 &  -0.011 \\
     222.147 &        0 & e &  2.5 & & 450153 & o &  2.5 & -1.77 &  2.29E+09 &   0.080 \\
     223.324 &        0 & e &  2.5 & & 447780 & o &  3.5 & -0.23 &  7.82E+10 &   0.593 \\
     223.527 &        0 & e &  2.5 & & 447373 & o &  1.5 & -1.09 &  1.08E+10 &   0.559 \\
\hline
\end{longtable}
}
}

\end{document}